\documentclass{ws-p8-50x6-00}

\RequirePackage{xspace}
\RequirePackage{subfigure}

\newcommand{\WW}{\ensuremath{\rm W^+ W^-}\xspace}
\newcommand{\rts}{\ensuremath{\sqrt{s}}\xspace}
\newcommand{\gev}{\ensuremath{~{\rm GeV}}\xspace}

\newcommand{\qqqq}{\ensuremath{q\bar{q}q\bar{q}}\xspace}
\newcommand{\nl}{\ensuremath{l\nu}\xspace}
\newcommand{\qq}{\ensuremath{q\bar{q}}\xspace}
\newcommand{\ql}{\ensuremath{q\bar{q}l\bar{\nu_l}}\xspace}
\newcommand{\mw}{\ensuremath{\rm M_W}\xspace}
\newcommand{\gw}{\ensuremath{\rm \Gamma_W}\xspace}

\begin{document}

%
%

\title{Colour Reconnection at LEP2}

\author{Raja Nandakumar}

\address{Tata Institute of Fundamental Research, India\\
         E-mail: Raja.Nandakumar@cern.ch}

\maketitle

%
%

\abstracts{Colour reconnection is the final state interaction
between quarks from different sources. It is not yet fully
understood and is a source of systematic error for W-boson mass
and width measurements in hadronic \WW decays at LEP2. The
methods of measuring this effect and the results of the 4 LEP
experiments at  $183\gev\leq\rts\leq 202\gev$ will be presented.}

%
%

\section{Introduction}

The W--boson is a carrier of charged weak interactions and it
decays into lepton--neutrino (\nl) or quark--antiquark (\qq) pairs.
The LEP2 W--factory\cite{LEP2PHY} at CERN, Geneva has produced thousands of
W--pair events which have been collected by the 4 LEP experiments
(Aleph, Delphi, L3 and Opal).

The typical distance between W decay vertices at LEP2 (0.1 fm) is
much smaller than the typical hadronisation distance (1 fm). When
both the W's decay in the \qq mode, this leads to the possibility
of gluon exchange between quarks from different W's, called
Colour Reconnection (CR).

\begin{figure}[hb]
  \vspace*{-7mm}
  \center
  \epsfig{file=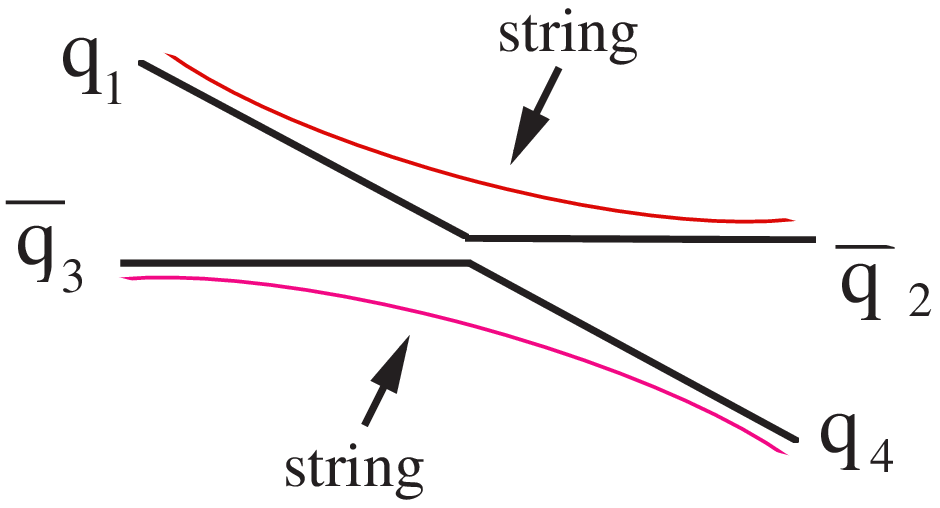,width=0.35\linewidth}\hspace{10mm}
  \epsfig{file=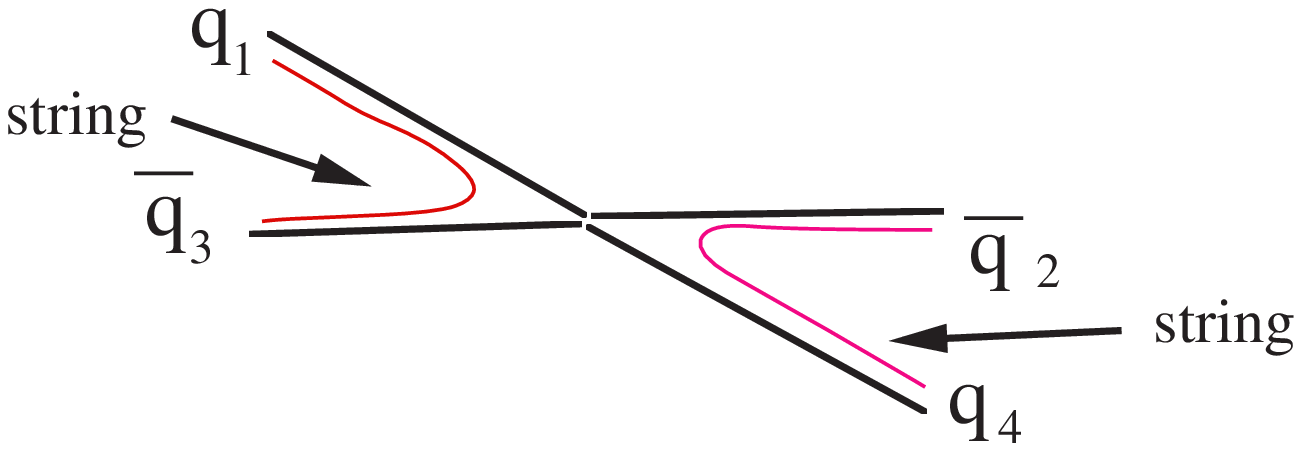,width=0.49\linewidth}

  No Reconnection \hspace*{20mm} Immediate Reconnection
  \caption{Illustration of Colour Reconnection (CR) between two W's
  decaying into $q_1\bar{q}_2$ and $q_3\bar{q}_4$, respectively.}
  \label{fig:strings}\vspace{-5mm}
\end{figure}

CR\cite{DELSHARKA} can thus be defined as the strong interaction between
independent colour singlets, before hadron formation. This causes a change
in the string configuration (figure \ref{fig:strings}) leading to changes
in the reconstructed 4--momentum of the W's. Thus, a study of CR tells
us about the space--time evolution of hadronic systems and about the
systematic uncertainty of the \mw and \gw measurements at LEP2.

%
%

\section{Methods of measuring CR}

Changes in the string configuration change the momentum distribution of
the particles and hence, also the expected event multiplicity. Thus, these
quantities carry information about CR. The typical analysis procedure to
study CR is to compare the data with Monte Carlo events simulated using various
models\cite{CRINMC} (SKI, SKII, GH, etc). We note that the \qq sector of
\ql events is a source of W's without CR.

In \qqqq events, there is an additional problem of pairing the jets
correctly, due to the existence of 3 different ways of pairing up 4 jets
into two W--bosons. This problem does not arise in the \ql events.

\section{Multiplicity based measurements}

\begin{figure}
  \vspace*{-8mm}
  \center \epsfig{file=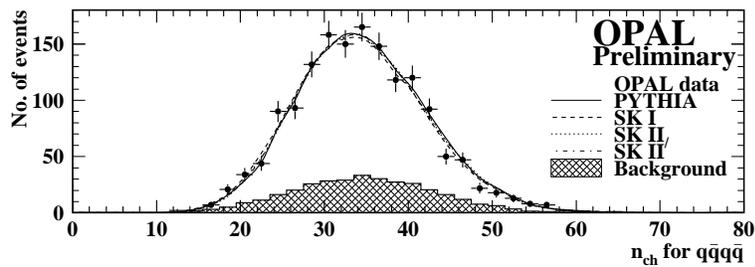,width=0.85\linewidth}\vspace{-2mm}
  \vspace{-3mm}
  \caption{Charged particle multiplicity}
  \label{fig:opal_nch}\vspace{-2mm}
\end{figure}

CR typically leads to a reduction in the string length. This implies
that the average multiplicity is decreased. This is a small effect, as
shown for charged particles in OPAL data (figure \ref{fig:opal_nch}).

One can define the average charged multiplicities
\begin{itemize}
  \vspace{-2mm}
  \item[] $\langle N_{\rm ch}^{4q}\rangle$ : from \qqqq events,
  \hspace{8mm} $\langle N_{\rm ch}^{2q}\rangle$ : from \qq system of \ql events
  \vspace{-3mm}
  \item[] \hspace{10mm} $\Delta\langle N_{\rm ch}\rangle =
    \langle N_{\rm ch}^{4q}\rangle - 2\langle N_{\rm ch}^{2q}\rangle$
    \hspace{10mm}($\neq$ 0 for CR) \\
  \vspace{-6mm}
\end{itemize}
Averaging over all LEP data \cite{CRMULT}, the author obtains for
$\Delta\langle N_{\rm ch}\rangle$,
a value of about $0.22 \pm 0.20 \pm 0.35$. This result is consistent with 0
and the error is dominated by systematics (mainly fragmentation modelling).

Another approach\cite{CRMULT} is to study the fragmentation functions
$x_p = 2p / \rts $ and $\xi = -log{x_p}$, where $p$ is the momentum of
the particle under consideration and $\rts$ is the center of mass energy.

\begin{figure}
  \vspace*{-3mm}
  \center \epsfig{file=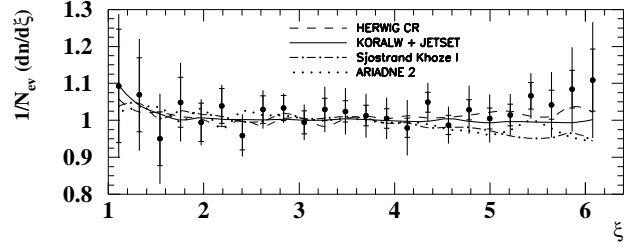,width=0.7\linewidth}\vspace{-2mm}
  \caption{Sensitivity of the fragmentation function to CR (Aleph)}
  \label{fig:aleph_xi}\vspace{-2mm}
\end{figure}

The sensitivity of the fragmentation function to the presence of CR
is shown in figure \ref{fig:aleph_xi}. Sensitivity is maximum in the region
$4.5 \le\xi\le 6.1 ~~\Rightarrow~~ 0.001 \le x_p\le 0.005$. Similar studies
can also be performed with heavy hadrons (eg. protons, kaons) which will
be more sensitive to CR effects. However, the net result is that the
systematic error of the method is greater than the expected effect and
no significant non-zero result is seen in the data.

\section{Particle flow based measurements}

\begin{figure} \center
  \vspace*{-4mm}
  \epsfig{file=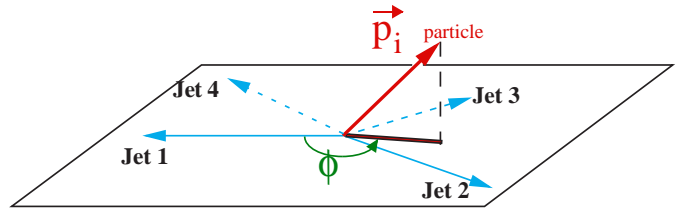,width=0.75\linewidth}
  \caption{The particle flow method}
  \label{fig:partflow_3}
\end{figure}

\begin{figure} \center
  \subfigure[The raw distribution]{
    \epsfig{file=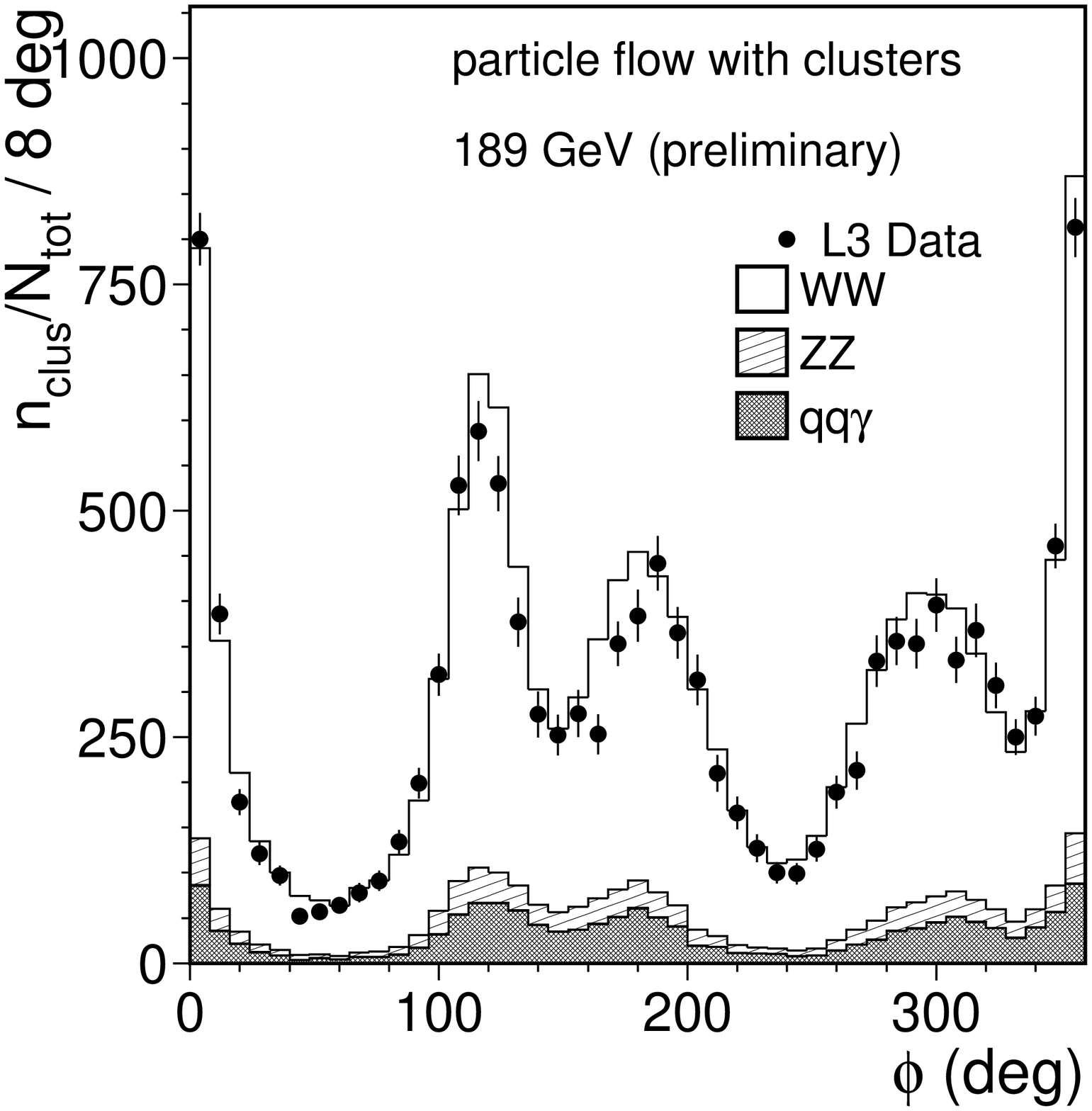,width=0.47\linewidth}
    \label{fig:l3_pf1}
  }
  \subfigure[The symmetrised distribution]{
    \epsfig{file=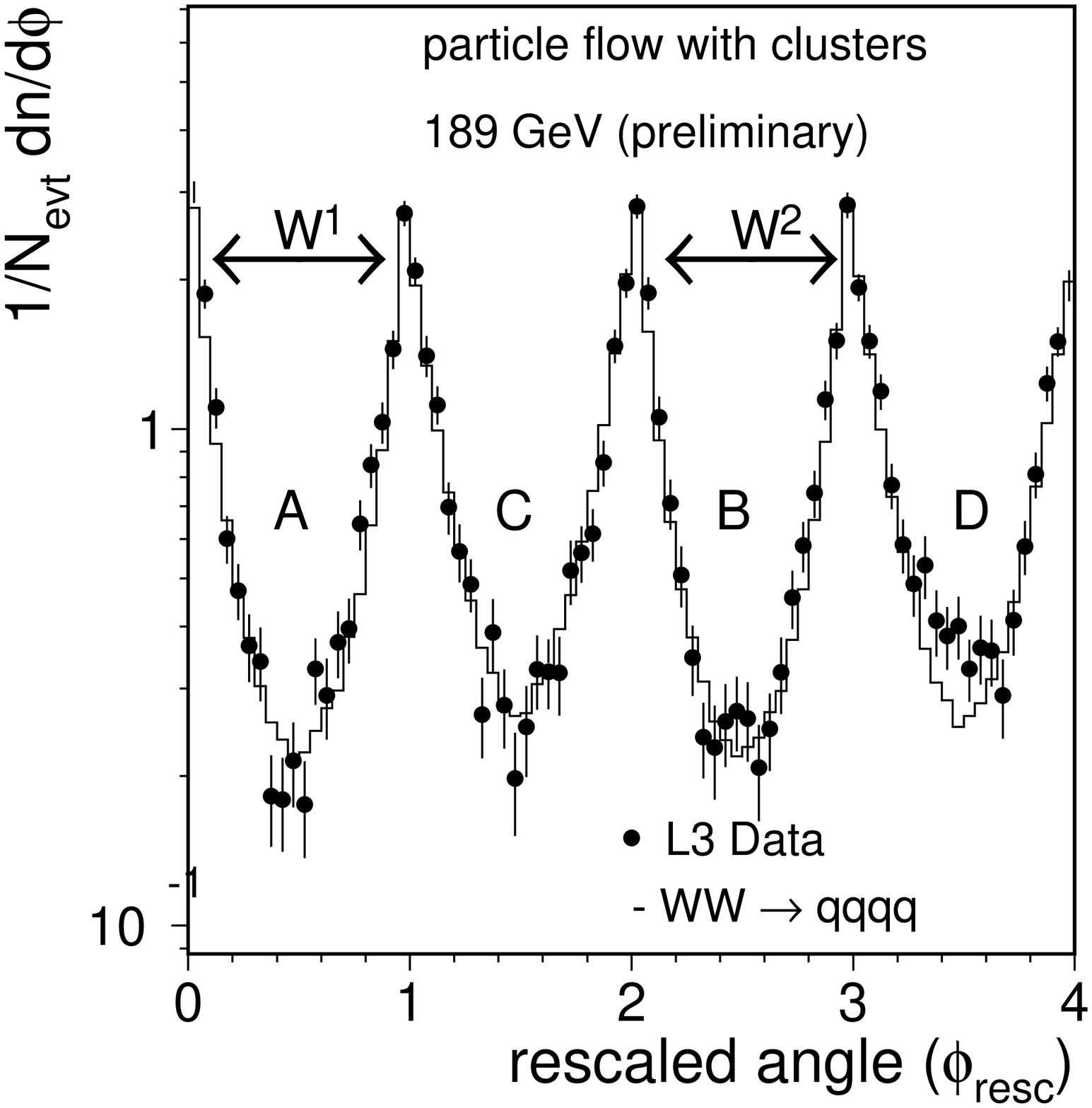,width=0.45\linewidth}
    \label{fig:l3_pf3}
  }
  \caption{The particle flow distributions from L3}
\end{figure}

The particle flow method\cite{CRPART} studies the event's string topology by
looking at the particle momentum distribution across the jets. The event
(satisfying the topology in figure \ref{fig:strings}) is forced into 4 jets
to reconstruct the 4 parent quarks. Two adjacent jets -- the most energetic jet
(jet 1) and the jet associated to it (jet 2) -- are used to form a
plane and the particles in the event are projected onto this plane. The
angle of the projection ($\phi$) from jet 1 is called the particle flow of
the event (figure \ref{fig:partflow_3}). The
distribution for $\phi$ is shown in the figure \ref{fig:l3_pf1}.

The angle $\phi$ is rescaled to the jet--jet angle to obtain
$\phi_{\rm resc} = \frac{\phi}{\phi_{jj}}$. The distributions are symmetrised
between the 3 adjacent pairs of jets,
as shown in figure \ref{fig:l3_pf3}. The energy flow distributions
correspond to the same distributions, with each particle in the distribution
being weighted with its energy, normalised to the visible energy in the event.
The background is subtracted from the data distribution before comparison
with Monte Carlo.

\begin{figure} \center
  \epsfig{file=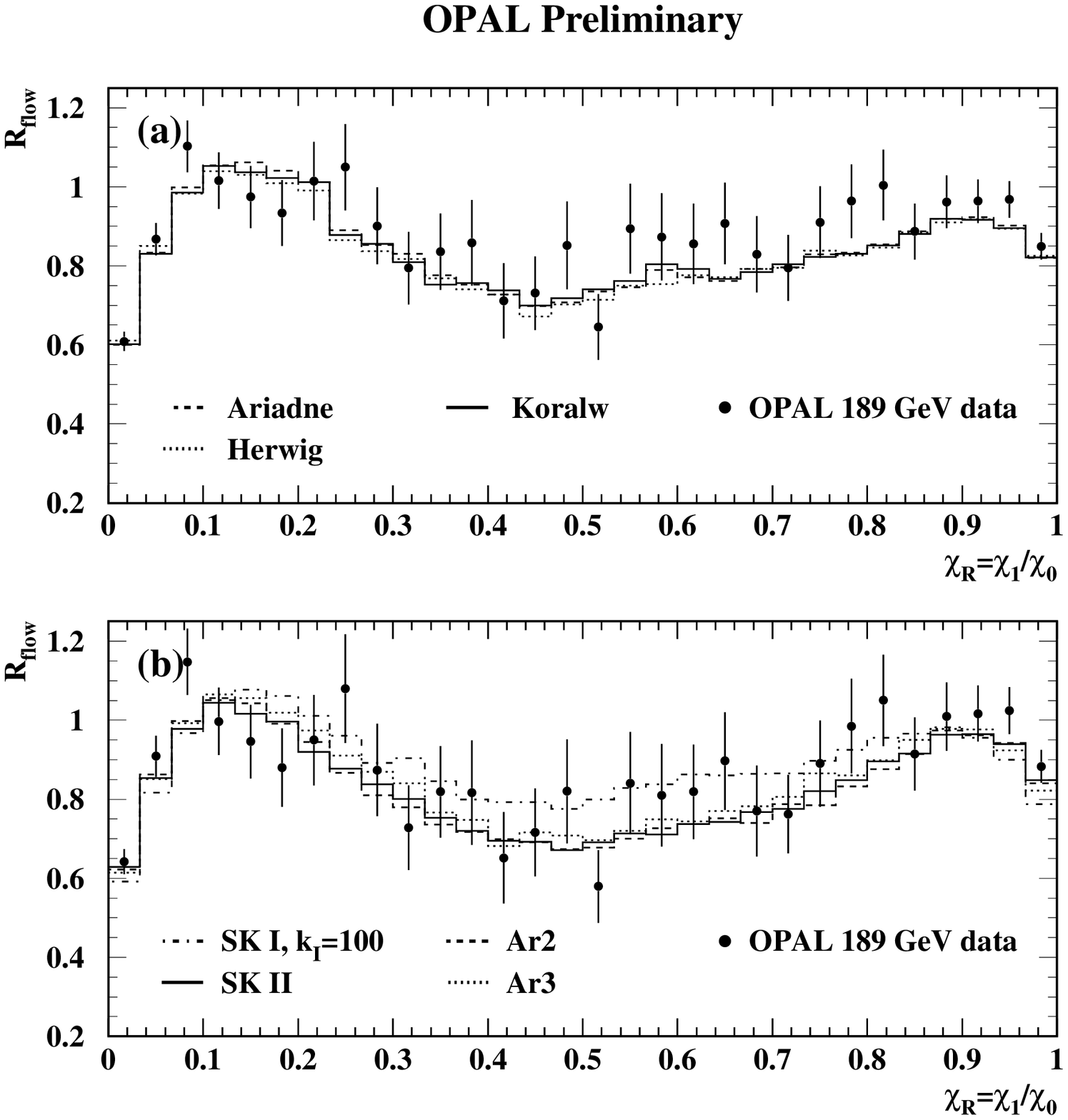,width=0.49\linewidth}
  \epsfig{file=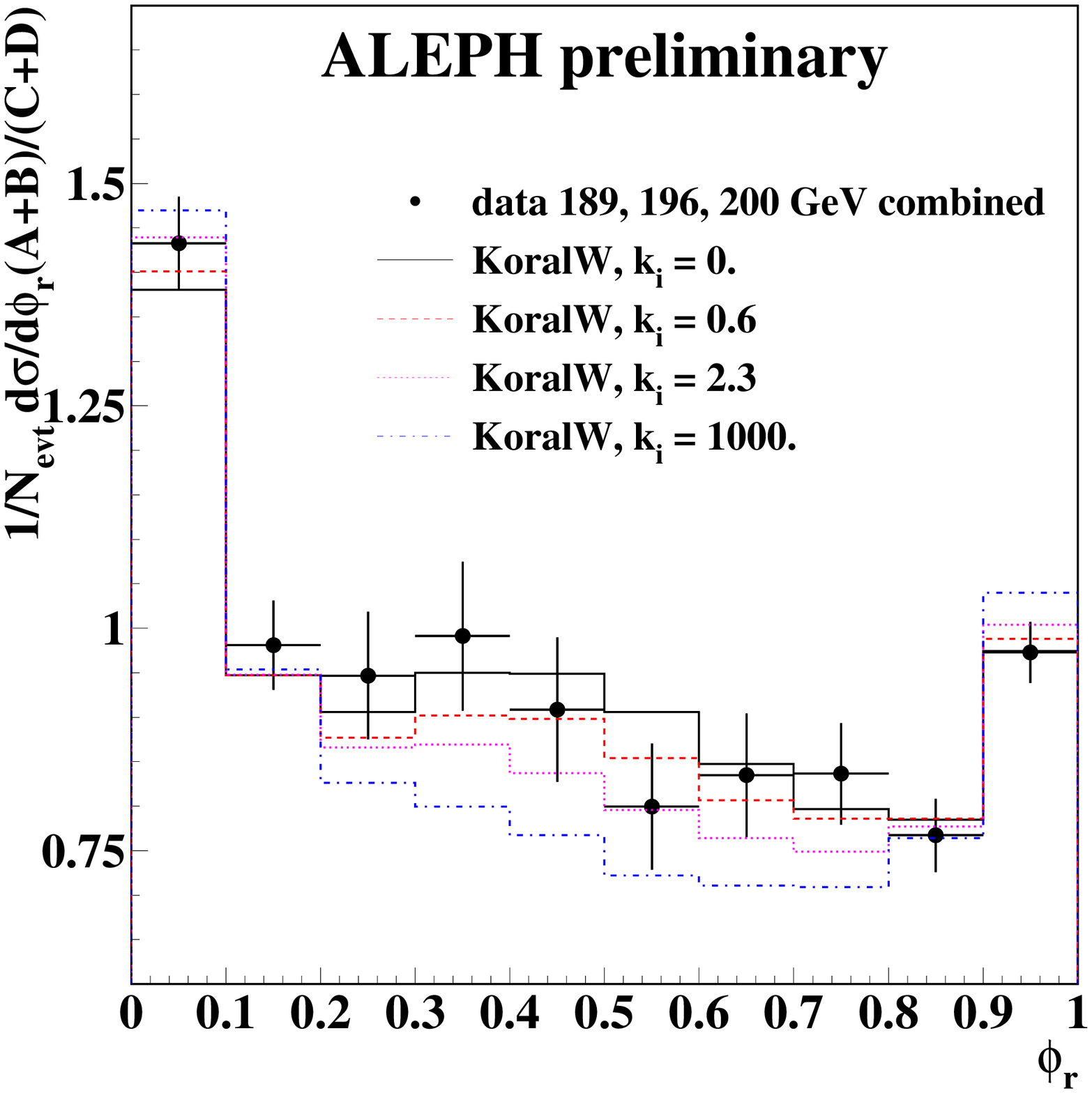,width=0.45\linewidth}
  \caption{Distributions for $\rm\frac{A+B}{C+D}$}
  \label{fig:l3_pf56}
\end{figure}

The regions A and B in figure \ref{fig:l3_pf3} arise when the two jets
come from the same W and the regions C and D arise when the two jets are
wrongly paired. Hence, one expects CR to cause a depletion in A and B,
and an enhancement in C and D. This can be clearly seen by taking the ratio
$\rm\frac{A+B}{C+D}$ as shown in figure \ref{fig:l3_pf56}.
This ratio also has the advantage of cancellation of some systematics.

The distributions shown in figure \ref{fig:l3_pf56} are integrated over
their most sensitive region, to obtain
$R = \int_{0.3}^{0.7}\frac{A+B}{C+D}~{\rm d}\phi_{resc}$. This $R$ is a good
estimator of the presence of CR in the data. As seen from the L3 results
in table \ref{tab:L3part} for 189 GeV, the systematics for this method are
much smaller than the statistical error and the data are inconsistent with
the result of no CR at the $2\sigma$ level.

For this method to work, one would obviously need jets that are clearly
identified and with a high probability of correct pairing. This leads to
a low efficiency ($\simeq 15 \%$) in selection of the WW events.

\begin{figure}
  \epsfig{file=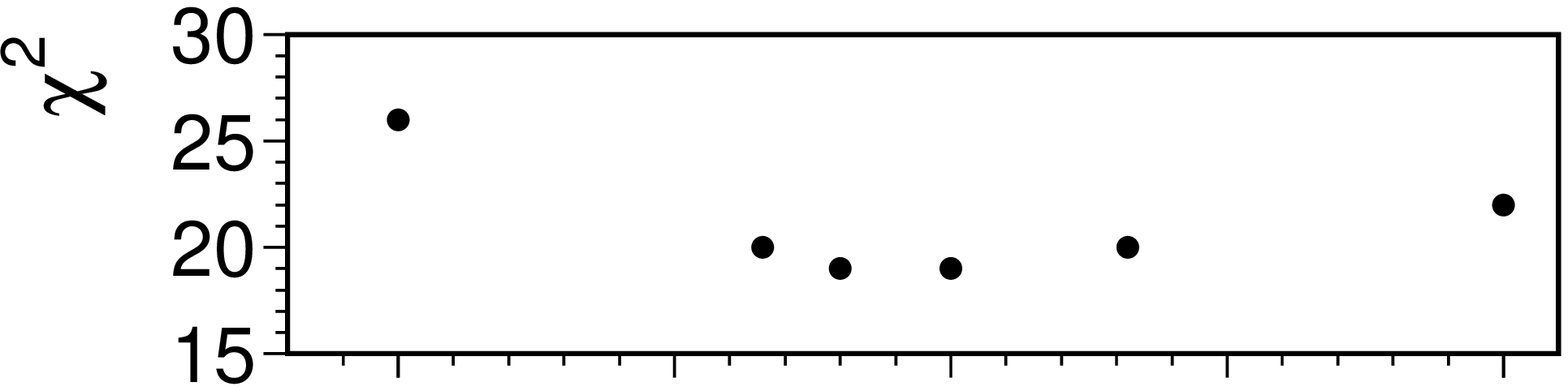,width=0.49\linewidth}\\
  \epsfig{file=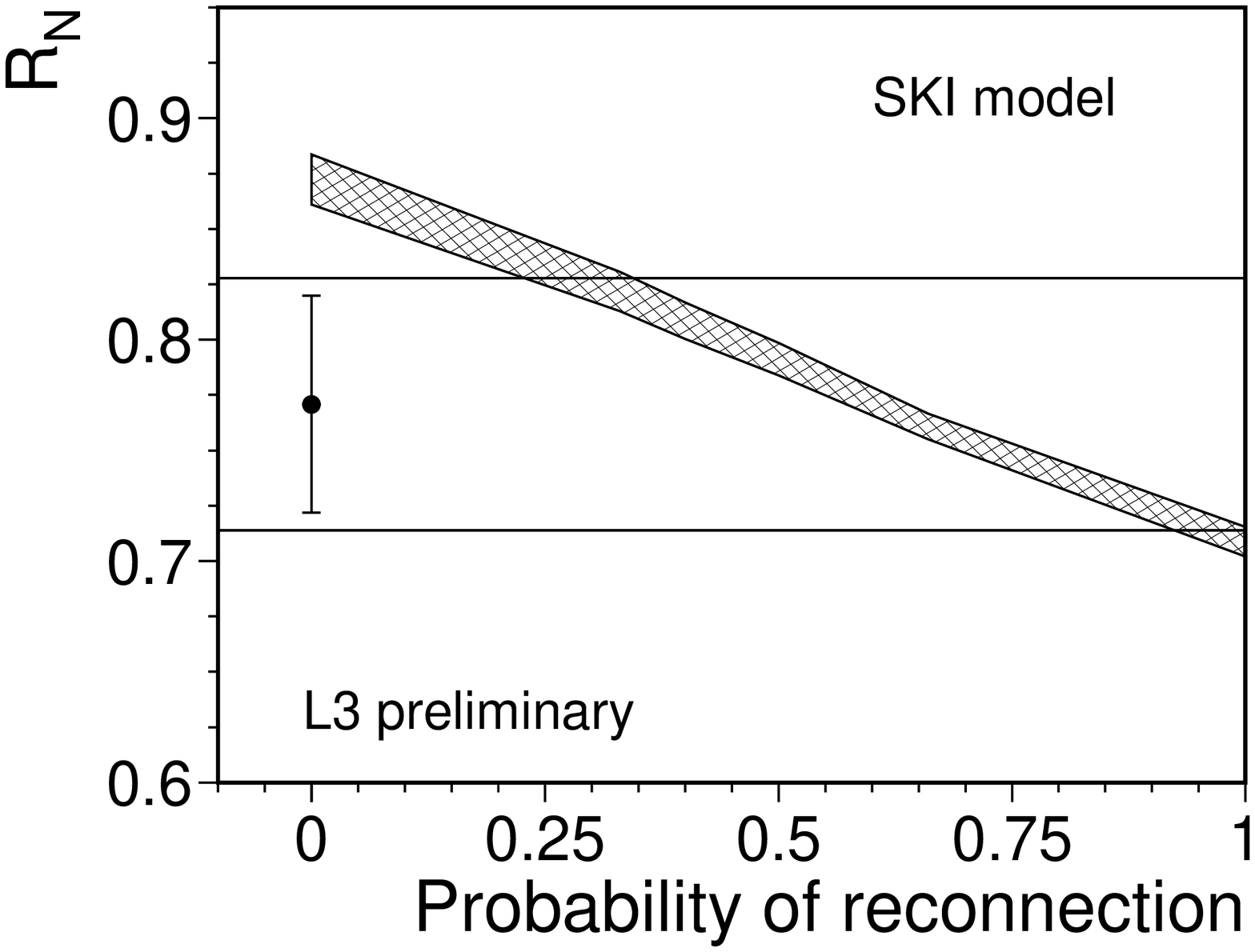,width=0.49\linewidth}

  \vspace*{-0.52\linewidth}\hfill
  \epsfig{file=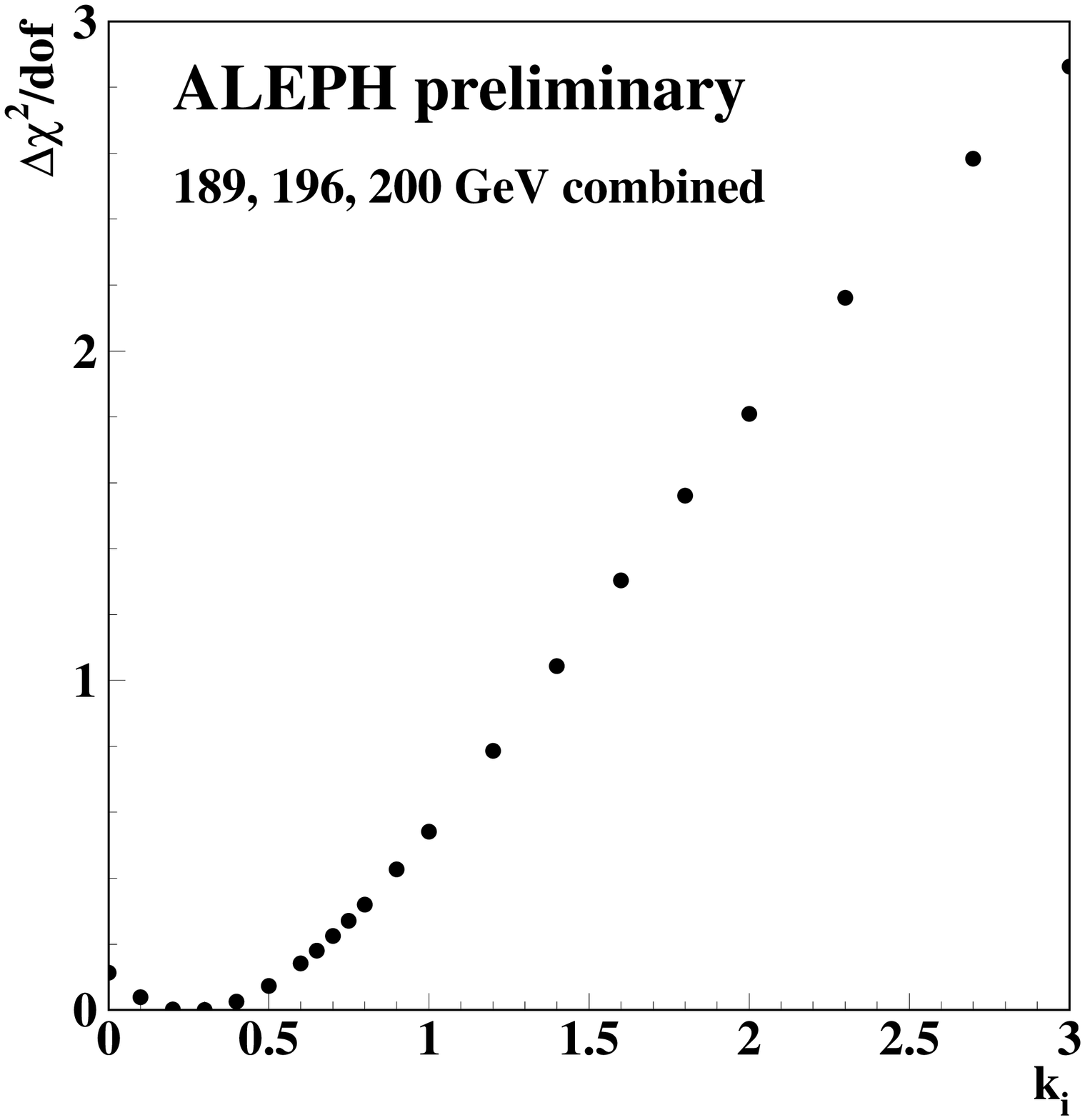,width=0.49\linewidth}

  \caption{Chisquared vs SKI CR probability for L3 and Aleph data}
  \label{fig:chisq}
\end{figure}

Similar analyses have been performed by Aleph and Opal. One can define a $\chi^2$
to estimate the best fit of data to Monte Carlo for different values of CR
probability (figure \ref{fig:chisq}). This is done using the SKI model, where
the parameter $\rm k$ is proportional to the probability of CR. The data from
Aleph, L3 and Opal, prefers in general, the existence of CR
with probabilities ranging from 15\% to 60\% with an error from each
collaboration, of about 30\%. A combination of the results will result in an
improvement of numbers.

\section{Summary}

CR can be measured in hadronic WW decays at LEP. The multiplicity and
fragmentation function based methods do not yield significant results, and
none are expected due to high systematic errors inherent to the method.
However, first indications of CR have been seen using the particle flow method.

The W--boson mass and width measurements are major aims for LEP2. Current
estimates of systematic uncertainties from CR for the mass is 50 MeV,
fully correlated across experiments, while it varies from 40 to 70 MeV
for the width. It is hoped that further studies on CR, especially through
the particle flow method will help in improving these numbers.

\section*{Acknowledgement}

The author would like to thank TIFR, CERN and KFKI for their support during
ISMD 2000. Many thanks too to Som Ganguli, Martin, Paul, Luca, John, Dominique,
Gagan and the LEP collaborations for their kind help during preparations.

%
%

%
%

\begin{table}[hb]
  \center
  \begin{tabular}{|c|c|c|}
    \hline
    $R$ & Data & Monte Carlo : No CR \\
    \hline
    Particle flow  & $0.771\pm 0.049\pm 0.029$ & $0.868\pm 0.007$ \\
    Energy flow    & $0.593\pm 0.058\pm 0.020$ & $0.696\pm 0.009$ \\
    \hline
  \end{tabular}
  \caption{189 GeV results from L3 for particle flow}
  \label{tab:L3part}
\end{table}

%
%

%
%


\begin{thebibliography}{99}

\bibitem{LEP2PHY} Physics at LEP-II, CERN 96--01, Vol.1

\bibitem{DELSHARKA} \v{S}. Todorova-Nov\'a, DELPHI 96--158 PHYS 651

\bibitem{CRINMC}
  T. Sj\"ostrand and V. Khoze, Z. Phys. C 62 (1994) 281 \\
  T. Sj\"ostrand and V. Khoze, E. Phys. J. C 6 (1999) 271 \\
  G. Gustafson and J. H\"akkinen, Z. Phys. C 64 (1994) 659

\bibitem{CRMULT}
  ALEPH Collaboration, Contr. to ICHEP 2000, ALEPH 2000 -- 058 \\
  DELPHI Collaboration, CERN-EP/2000-023 \\
  L3 Collaboration, Contr. to ICHEP 2000 \\
  OPAL Collaboration, CERN EP / 98 -- 196 ; Physics note, PN 412

\bibitem{CRPART}
  ALEPH Collaboration, Contr. to ICHEP 2000 \\
  L3 Collaboration, Contr. to ICHEP 2000 \\
  OPAL Collaboration, Physics note, PN 412
  

\end{thebibliography}
\end{document}